\documentclass[aps,pra,10pt,superscriptaddress, preprint]{revtex4-2}
\usepackage{graphicx}
\usepackage{amsmath,amsthm,amssymb,dsfont}
\usepackage{mdframed}
\usepackage{orcidlink}

\usepackage{hyperref}
\usepackage{soul}

\begin{document}

\title{Reply to ``Counterfactual communication not achieved yet – A Comment on Salih et al. (2022)''}
\author{Hatim Salih}
\affiliation{School of Physics, Engineering and Technology
University of York, York YO10 5DD, United Kingdom}
\author{Jonte R. Hance\,\orcidlink{0000-0001-8587-7618}}
\email{jonte.hance@newcastle.ac.uk}
\affiliation{School of Computing, Newcastle University, 1 Science Square, Newcastle upon Tyne, NE4 5TG, UK}
\affiliation{Quantum Engineering Technology Laboratories, Department of Electrical and Electronic Engineering, University of Bristol, Woodland Road, Bristol, BS8 1US, UK}
\author{Will McCutcheon}
\affiliation{Institute of Photonics and Quantum Science, School of Engineering and Physical Sciences, Heriot-Watt University, Edinburgh, EH14 4AS, UK}
\affiliation{Quantum Engineering Technology Laboratories, Department of Electrical and Electronic Engineering, University of Bristol, Woodland Road, Bristol, BS8 1US, UK}
\author{John Rarity}
\email{john.rarity@bristol.ac.uk}
\affiliation{Quantum Engineering Technology Laboratories, Department of Electrical and Electronic Engineering, University of Bristol, Woodland Road, Bristol, BS8 1US, UK}

\begin{abstract}
In his Comment on our recent paper “The laws of physics do not prohibit counterfactual communication”, \textit{npj Quantum Information} (2022) 8:60, Popescu argues that the claims of the paper are invalid. Here, we refute his argument, showing that it is based on {ignoring the specifics of what we set out to prove (that counterfactual communication is possible \emph{for post-selected particles}, {and more specifically in these cases is not prohibited by the weak trace or consistent histories criteria for particle path)}), followed by an} unwarranted simplification of the protocol. Moreover, the Comment’s excursion into interpretation is misplaced. Our communication protocol is a precisely defined one that allows two remote parties, albeit rarely, to communicate an arbitrarily long binary message, with arbitrarily high accuracy. This is not a matter of interpretation---as the concrete example given in our paper in question illustrates. As for our overarching claim that no particles are exchanged in the course of this communication, we have already demonstrated this both theoretically and experimentally, {in the postselected case we consider}, {as per the weak trace and consistent histories criteria for path of a quantum particle}.
\end{abstract}

\maketitle

{The simplest issue with Popescu's comment \cite{Popescu2025CommentLaws} is that it mistakenly infers that we are saying our protocol \cite{salih2022laws} is counterfactual \emph{even for failed runs} (those where no information is transferred from Bob to Alice, and the photon ends up in detector $D_3$). This is patently wrong---in our abstract, and throughout the body of our paper, we repeatedly state we are talking about the postselected case (when Alice's photon arrives either in her detector $D_0$ or her detector $D_1)$. At the beginning of our Results Section, we even explicitly say that one of the two aims of the protocol we construct in our paper construct one where}

{\textit{``It can be shown unambiguously that Alice’s \textbf{post-selected} photons have never been to Bob.''}}

{Popescu's Comment instead treats it as though we are referring to counterfactuality even outside this postselected case.} {Further, the toy protocol in our paper is one modified from Salih et al's 2013 protocol \mbox{\cite{Salih2013Protocol}} specifically designed to show that the weak trace \mbox{\cite{Vaidman2013Past}} and consistent histories \mbox{\cite{Griffiths2002ConsistentTheory}} criteria for particle path do not prohibit counterfactual communication, given these are the criteria on which criticisms of Salih et al's original protocol had been based at the time of publication (see e.g., \mbox{\cite{Vaidman2015, Vaidman2016CommentState,Vaidman2014a, Wander2021Analyzing, Griffiths2016}}. While our toy protocol is formed of one outer interferometer cycle of Salih et al's 2013 protocol, there is a reason this protocol was originally give for two or more outer cycles---this one-outer-cycle toy protocol can give erroneous ``0-bit'' clicks, while in two-or-more outer-cycle versions of Salih et al's protocol, this is not the case---whenever Alice receives a bit, it is exactly the same bit-value Bob intended to send. While we reply here to Popescu's criticisms of the toy protocol, we wish to make clear the toy protocol is designed specifically so we could show that counterfactual communication is not prohibited by the weak trace and consistent histories approaches, given the complexity of analysis with these tools, and so lacks features (e.g., interference between inner and outer path of outer interferometer) which Salih et al's 2013 protocol possesses, which more easily rebut Popescu's criticisms. (Note also that, while we show in \mbox{\cite{salih2022laws}} that counterfactual communication is possible by the weak trace and consistent histories criteria of particle path, this does not mean we advocate these as criteria for the path of a quantum particle---indeed, \mbox{\cite{Salih2023Wormholes}} challenges the consistent histories criterion, and \mbox{\cite{Hance2023Weak}} challenges the weak trace criterion (with further discussion in \mbox{\cite{Vaidman2023CommentWeak,Hance2023ReplyWeak}}). One could quite easily come up with alternative definitions for particle presence, which would lead to different conclusions about what protocols are, and are not, ``counterfactual''. Similarly, even if you accept a given criteria for a quantum particle being in a certain location, a protocol being counterfactual by said criteria does not rule out there being indirect influences between Alice and Bob---indeed, one could argue that there must be {\em some} indirect influence/interference effect, otherwise it would be impossible to communicate using such a counterfactual protocol. \mbox{\cite{Hance2024CFBAIG}}, again by one of the authors, considers whether indirect influences in scenarios like this may be caused by back-action effects onto the system by the meter during measurement.}

We {next} note that the caption of Fig.~1 of our paper \mbox{\cite{salih2022laws}} indicates clearly that detector $D_0$ could have been placed immediately after the topmost beamsplitter---meaning that interference only takes place in the concatenated interferometers on the right. The author of the Comment takes several paragraphs to acknowledge this. The caption of Fig.~1 explains the reason for placing $D_0$ after the bottom-most beamsplitter: so that the setup exactly includes the equivalent of one outer cycle of Salih \textit{et al}'s Michelson-type (polarisation-based) protocol, laid-out sequentially in time \mbox{\cite{Salih2013Protocol}}. This allows the conclusions drawn for this single-outer-cycle protocol to be applicable to any of the outer cycles of {Salih et al's} 2013 counterfactual communication protocol, as does reference \mbox{\cite{Salih2023Wormholes}}. Perhaps not fully aware of how that 2013 (polarisation-based) counterfactual communication protocol works, the Comment disputes this applicability. The important point to note, however, is that when the inner interferometers are not blocked---i.e., bit `0', which is the case where counterfactuality had been questioned---no interference should take place in the outer interferometers, just as in the present single-outer-cycle protocol.

The Comment goes on to suggest a threefold simplification of our Protocol. First, the Comment suggests that the topmost polarising beam-splitter PBS and the polarisation rotator HWP1 can be removed, and replaced with a biased coin to decide whether to fire the photon into the concatenated interferometers on the right or not. This is correct—--and something that has already been shown in an ``After the Paper'' (on Springer Nature’s Physics Community) less-technical explanatory article on our paper in question \cite{BehindThePaper}.

The next simplification the Comment makes is to remove the coin toss altogether. It says:

\textit{``Then even the coin toss is not necessary. If Alice and Bob want to perform the original protocol for $N$ runs with probability p of live runs, they can simply use the basic apparatus for $pN$ runs, then turn it ``off'' and go home both, with Alice guessing ``no blocking'' for the rest of the $(1-p)N$ runs, while Bob doesn't even have to attempt transmitting anything.''}

When referring to $N$ runs, does the Comment mean one run per bit choice by Bob? In this case, having got rid of our use of the coin, the space of possible $N$-bit whole messages that can be accurately communicated with some non-zero probability is limited to essentially one $N$-bit message only: the message where all $pN \approx N$ bits have bit-value `1'. {In all other cases, the post-selection would fail (the photon would end at $D_3$), so the message would need to be disregarded.} Contrast this with our protocol, where any $N$-bit string from Bob can be entirely and accurately communicated, with some non-zero probability (that goes down as accuracy is set higher). Further, in the above simplification, `0' guesses by Alice, corresponding to ``no blocking'', are completely uncorrelated to Bob's bit choices. Contrast this with our protocol, where Alice's registered bits {in the postselected case}, whether `0' or `1', are as correlated with Bob's bits as desired. {We can see this by noting that, in our protocol, while each run of the protocol could only ever have transmitted possible one message, these messages are different each run, given the probabilistic nature of whether the photon goes into the inner interferometer chain or the outer path---and the protocol fails until we reach a run where the allowed message aligns with Bob’s intended message. This is different to Popescu’s protocol, where, for every run, the only message which could be sent that run is the one where “where all $pN \approx N$ bits have bit-value 1.” Therefore, unlike in Popescu’s protocol, communication is possible through our protocol.}

Alternatively, when referring to $N$ runs, does the Comment mean $N$ runs per bit? In that case, Alice gets to compare the frequency of her `0' guesses to how often $D_1$ clicks (which indicates `1'), turning a blind eye to information-carrying photons crossing between Alice and Bob during failed runs. {However, the ``blind eye'' here is different to the postselection we consider (where any failed run leads to the entire data-set being thrown away), and so} again, this is different from our protocol, where there is one run per bit choice by Bob. This is clear from the definition of the protocol at the beginning of the Discussion section \cite{salih2022laws}, 

\textit{``In each round of the proposed experiment Bob chooses a bit, $X$, he would like to communicate to Alice. He blocks (does not block) his channel when $X=0$ ($X=1$). Alice then prepares a single photon, passes it through the system and it is either detected in one of the detectors $D_0$, $D_1$ and $D_3$, or is lost to Bob’s blocking device. If Alice detects the photon in either $D_0$ or $D_1$, then the round was successful and Alice assigns the estimated values $X_{est}=0$ and $X_{est}=1$ to detections in $D_0$ and $D_1$ respectively. If the round was not successful another round is performed until she obtains a successful outcome.''}

That the Comment's simplification corresponds to a different protocol than ours is also clear from the concrete example in our paper's Discussion section, of Bob sending a 16-bit string, where the whole message is discarded anytime Alice fails to register even a single bit {(i.e., any time any photon in the string goes to Bob's blocker, or $D_3$)}. Thus, contrary to what the Comment imagines, no bit-information whatsoever is gained from failed runs where neither of Alice's detectors clicks. Now, given a big enough stack of such messages, eventually, Bob would be able to communicate a 16-bit message. Communication accuracy can be made as high as desired, albeit at the expense of more messages being discarded: demonstrating that the laws of physics do not prohibit counterfactual communication. Why is this a stronger result than previous counterfactual protocols? As the paper makes clear from the outset, the fact that no photons have crossed the channel in such a scenario can be verified using, not only weak measurements, but also consistent histories \cite{salih2022laws}.

Finally, building on the above {over}simplification, the Comment goes on to make a further erroneous one, namely that one need only focus on the standard interaction free measurement setup. By now the Comment is outright considering a different protocol, which makes the final point about post-selection not relevant either.

It could be argued that the postselection itself is not counterfactual, given Alice knows in cases where she doesn't receive a photon at either of her detectors that the photon has gone to Bob, and either been absorbed by his blockers if he blocked, or has gone to detector D3 if he didn't block (or has been lost in the apparatus). These cases could be said to be responsible for the information transfer observed by Alice in her postselected cases. However, assuming Alice sticks to our protocol, and only infers information from D0 or D1 clicking, then in any case she is directly inferring information, the photon will (by the weak trace and consistent histories conditions at least) not have been to Bob. Therefore, so long as Alice throws out any messages where, for any bit, she doesn't receive a photon, then, for that entire message, no photons will have been to Bob, and so that message (at least by these conditions) is counterfactual. Unlike the case Popescu gives, where there is only one message for which this is possible, for our protocol (given postselection, which, as admitted, could be considered non-counterfactual), any message Bob chooses could be sent to Alice this way.

In conclusion, while the protocol presented in our paper in question \cite{salih2022laws} can be simplified without violating counterfactuality---as has been shown previously \cite{BehindThePaper}---any `simplification' that either violates counterfactuality {(as Popescu's does, by not considering our postselection criterion)} or results in markedly different statistics, is by definition no longer our protocol, as is the case with the simplification presented in the Comment. We note that more efficient counterfactual communication protocols, including the first such protocol \cite{Salih2013Protocol}, use more than one outer cycle (comprising the big diamond shape) whereby interference effects in outer cycles are employed. In our paper in question, by contrast, we prove experimentally using a setup consisting of a single outer cycle that the laws of physics {(or at least the weak trace and consistent histories criteria for the path of a quantum particle)} do not prohibit counterfactual communication. Building on this, the counterfactuality of more complex schemes can be better understood and supported \cite{Salih2023Wormholes}---including counterfactual qubit transport (or counterportation), counterfactual ghost imaging, and exchange-free quantum computation \cite{Salih2016Qubit,Salih2021EFQubit,Hance2021CFGI,Salih2023Wormholes}.

\textit{Acknowledgements—}HS acknowledges support from the University of York. JRH acknowledges support from Royal Society Research Grant RG/R1/251590, Hiroshima University's Phoenix Postdoctoral Fellowship for Research, and the University of York's EPSRC DTP grant EP/R513386/1. JGR and JRH acknowledge support from Quantum Communications Hub funded by EPSRC grants EP/M013472/1 and EP/T001011/1.

\textit{Competing Interests---} The authors declare that there are no competing interests.

\textit{Author Contributions---} All authors contributed equally to this article.

\end{document}